\documentclass[pdftex,twocolumn,epjc3]{svjour3}  
\smartqed  
\RequirePackage{graphicx}
\RequirePackage{amssymb}
\RequirePackage{amsmath}
\RequirePackage{bm}
\RequirePackage{mathptmx}      
\RequirePackage{flushend}
\RequirePackage{cuted}
\RequirePackage[numbers,sort&compress]{natbib}
\RequirePackage[colorlinks,citecolor=blue,urlcolor=blue,linkcolor=blue]{hyperref}
\newcommand{\be}[1]{\begin{equation}\label{#1}}
\newcommand{\ee}{\end{equation}}
\newcommand{\ba}[1]{\begin{eqnarray}\label{#1}}
\newcommand{\ea}{\end{eqnarray}}
\newcommand{\rf}[1]{(\ref{#1})}
\newcommand{\nn}{\nonumber}

\newcommand{\ov}{\bar}

\journalname{Eur. Phys. J. C}

\begin{document}\sloppy

\title{Scalar and vector perturbations in a universe with nonlinear perfect fluid
}


\author{Ezgi Canay\thanksref{e1,addr1}
        \and
        Ruslan Brilenkov\thanksref{e2,addr2}
        \and
        Maxim Eingorn\thanksref{e3,addr3}
        \and
        A. Sava\c{s} Arapo\u{g}lu\thanksref{e4,addr1}\and\\        
        Alexander Zhuk\thanksref{e5,addr5}}
\thankstext{e1}{e-mail: ezgicanay@itu.edu.tr}
\thankstext{e2}{e-mail: brilenkov@astro.rug.nl}
\thankstext{e3}{e-mail: maxim.eingorn@gmail.com}
\thankstext{e4}{e-mail: arapoglu@itu.edu.tr}
\thankstext{e5}{e-mail: ai.zhuk2@gmail.com}

\institute{Department of Physics, Istanbul Technical University, 34469 Maslak, Istanbul, Turkey \label{addr1}
           \and
          Kapteyn Astronomical Institute, University of Groningen, P.O. Box 800, 9700 AV Groningen, The Netherlands \label{addr2}
           \and
            Department of Mathematics and Physics, North Carolina Central University,
	1801 Fayetteville St., Durham, North Carolina 27707, U.S.A. \label{addr3}
           \and
            Astronomical Observatory, Odessa I.I. Mechnikov National University, Dvoryanskaya St. 2, Odessa 65082, Ukraine \label{addr5}                   
}

\date{Received: date / Accepted: date}

\maketitle

\begin{abstract}
We study a three-component universe filled with dust-like matter in the form of discrete inhomogeneities
(e.g., galaxies) and perfect fluids characterized by linear and nonlinear equations of state. Within the cosmic screening approach, we develop the theory of scalar and vector perturbations. None of the energy density contrasts associated with the distinct components is treated as small. Consequently, the derived equations are valid at both sub- and super-horizon scales and enable simulations for a variety of cosmological models.
\end{abstract}

\section{Introduction}
\label{sec:1}
Unraveling the physics behind the accelerated expansion of the late universe is one of the greatest challenges in modern cosmology. Going through a number of models that have been proposed to explain this phenomenon, one may first refer to Einstein's legendary cosmological constant as the underlying energy component. However, though it provides the sought-for behavior, it also brings along the question of what could be the origin of such a constant with the value required for the observed acceleration, yet unanswered. Perfect fluids with the linear equation of state (EoS) $p=\omega\varepsilon$, \mbox{$\omega=\mathrm{const}$} are also known to drive the acceleration provided that \mbox{$\omega <-1/3$}. Nevertheless, $\omega $ should be very close to $-1$ \cite{Planck2018} to be in agreement with observations. Alternatively, there exist the so-called quintessence \cite{quint1,quint2} models in which the corresponding source is a scalar field. These include cases leading to constant $\omega$ \cite{WL}, shown to impose severe restrictions on the form of the scalar field potential \cite{ZBG,zhuk1996}. Perfect fluids with the nonlinear EoS $p=f(\varepsilon)$ represent yet another class of models in this direction, and the Chaplygin gas model \cite{Chap1,Chap2,Chap3,Chap4} may be studied as an example of those. 

As one can see, there exists an extensive zoo of models that can explain the accelerated expansion of the universe. Obviously, the models to survive are those with predictions closest to the observational data. Among such criteria is the observed large-scale structure of the universe described on the basis of the widely studied perturbation theory \cite{Bardeen,Peebles,Mukhanov,Mukhanov2,Durrer,Rubakov}. In this direction, a cosmic screening approach has been proposed in the paper \cite{Eingorn1} for the scalar and vector perturbations, with the distinctive feature that the gravitational potential satisfies a Helmholtz-type equation and not a Poisson-type one. Consequently, at large cosmological distances from individual sources, the potential undergoes an exponential cutoff. Matter sources in \cite{Eingorn1} have the form of discrete point-like masses, and
it is also important to emphasize that no assumptions are made regarding the smallness of the associated energy density contrast, ensuring the validity of the model both at sub- and super-horizon scales. This approach has been further developed in the papers \cite{Claus1,Claus2,Eingorn2,EinBril2,Duygu,Emrah,f(R),coslaw,EE,Alvina}. Particularly in \cite{Claus1,Claus2}, generalizations to the case of perfect fluids with the linear $p=\omega \varepsilon$ and nonlinear $p=f(\varepsilon)$ equations of state have been performed, i.e. aside from point-like masses, additional components of two distinct types of perfect fluids have been included in the matter sources. Obviously, the linear component has the background EoS $\ov p=\omega \ov\varepsilon$. Nevertheless, for the nonlinear component, it is possible to write $\ov p=f(\ov \varepsilon)$ only in the case of small fluctuations where the expansion $p =f(\ov\varepsilon)+(\partial f/\partial\varepsilon)_{\ov\varepsilon}\delta\varepsilon+(1/2)
(\partial^2 f/\partial\varepsilon^2)_{\ov\varepsilon}\delta\varepsilon^2 +\ldots$ 
works well. This point is disregarded in some papers (see e.g. \cite{err1,err2,err3}) whereas in \cite{true1,true2,true3,true4,true5}, for instance, it is clearly stated that the relation $\ov p=f(\ov\varepsilon)$ does not hold in general. 

In \cite{Claus1}, the authors considered the case where energy density fluctuations of the nonlinear perfect fluid are small quantities. Meanwhile, density contrasts of the pressureless matter and perfect fluid with linear EoS were set arbitrary. In the present article, we investigate all three matter components with arbitrary density contrasts. Therefore, 
the considered model applies both to small/astrophysical scales, where matter fluctuations are large, and to large/cosmological distances, where the density contrast is small. We develop the theory of scalar and vector perturbations for this model within the cosmic screening approach and obtain a system of equations which enables the cosmological simulation for arbitrary forms of the function $f(\varepsilon)$.

The paper is structured as follows. In Section 2, the basic equations are presented and the theory of scalar and vector perturbations is constructed for the considered model. The main results are summarized in Section 3. Appendix is devoted to showing that the auxiliary equations employed in the main proof are satisfied within the adopted accuracy for arbitrary density contrasts.

\section{Scalar and vector perturbations in the cosmic screening approach}
\label{sec:2}
We investigate a universe which contains perfect fluids with nonlinear EoS $p_J=f_J(\varepsilon_J) ,\; J=1,2,\ldots$ Particular cases include the pressureless perfect fluid with $p=0$ and perfect fluids with linear EoS $p=\omega \varepsilon$,  $\omega=\mathrm{const}.$ The space-averaged distribution (denoted by the overbar) of these components determines the dynamics of the homogeneous and isotropic universe described by the Friedmann equation
\be{2.1} \frac{3{\mathcal H}^2}{a^2}=\frac{3H^2}{c^2}=\kappa \overline{\varepsilon}=\kappa\left(\overline{\varepsilon}_M+\sum\limits_I\overline{\varepsilon}_I+
\sum\limits_J\overline{\varepsilon}_J\right)\, ,\ee
in which $a\left(\eta\right)$ denotes the scale factor, the Hubble parameter ${\mathcal H}\equiv a'/a\equiv (da/d\eta)/a$, $\eta$ is the conformal time and the constant $\kappa\equiv 8\pi G_N/c^4$ ($G_N$ and $c$ are the Newtonian gravitational constant and the speed of light, respectively). The total averaged energy density $\overline\varepsilon$ in the above equation has been split into its constituent parts with respect to their types of EoS: index ``M" corresponds to pressureless matter (continuous as well as discrete) and indexes ``I" and ``J" correspond to perfect fluids with linear and nonlinear EoS, respectively. Fluctuations in the energy densities generate metric perturbations which, in the following, will be studied in terms of their associated scalar and vector components. The perturbed metric in the first-order approximation and in the Poisson gauge reads
\ba{2.2} &&ds^2= \,\nn\\
&&a^2\left[\left(1 + 2\Phi\right)d\eta^2 + 2B_{\alpha}dx^{\alpha}d\eta -\left(1-
2\Phi\right)\delta_{\alpha\beta}dx^{\alpha}dx^{\beta}\right]\, .\,\nn\\ \ea
The only approximation in our approach is that the metric corrections $\Phi$ and $B_\alpha$ as well as the peculiar velocities \mbox{$\tilde {\bf v}\equiv\left(\tilde v^1,\tilde v^2,\tilde v^3\right)$}, $\tilde v^{\alpha}\equiv d x^{\alpha}/d\eta$ are considered small: $\Phi, B_{\alpha}$, \mbox{$\tilde v^{\alpha} \ll 1$}. On the other hand, the smallness of the energy density and pressure fluctuations is not demanded, i.e. the density and pressure contrasts may exceed unity: $\delta\varepsilon/\overline\varepsilon$,  \mbox{$\delta p/\overline p >1$}. This serves as an indicator that our model works both on astrophysical and cosmological scales. Potentials $\Phi$ and \mbox{$\mathbf B\equiv\left(B_1,B_2,B_3\right)$} satisfy the linearized Einstein equations \cite{Claus1}
\ba{2.3} &&\triangle\Phi-3{\mathcal H}(\Phi'+{\mathcal H}\Phi)\,\nn\\
&=&\frac{1}{2}\kappa
a^2\left(\delta\varepsilon_M +\sum\limits_I\delta{\varepsilon}_I+\sum\limits_J\delta{\varepsilon}_J\right) ,\nn\\ \ea
\ba{2.4} &&\frac{1}{4}\triangle{B}_{\alpha}+\frac{\partial}{\partial x^{\alpha}}(\Phi'+{\mathcal H}\Phi)\,\nn\\
&=&\frac{1}{2}\kappa
a^2\left(-\frac{c^2}{a^3}\sum\limits_n\rho_n\tilde v^{\alpha}_n+\frac{\overline\rho_M c^2}{a^3}B_{\alpha}-\sum\limits_I(\varepsilon_I+p_I)\tilde
v_I^{\alpha}\right.\nn\\
&+&\left.\sum\limits_I(\overline\varepsilon_I+\overline p_I)B_{\alpha}-\sum\limits_J(\varepsilon_J+ p_J)\tilde v_J^{\alpha}+
\sum\limits_J(\overline\varepsilon_J+\overline p_J)B_{\alpha}\right)\, ,
\ea
where $\triangle$ is the Laplace operator in flat comoving space. In the Poisson gauge, the potential $\mathbf{B}$ is subject to the transverse gauge condition $\nabla\mathbf{B}\equiv \delta^{\alpha\beta}\partial B_{\alpha}/\partial x^{\beta}=0$. It should be noted that the indices of three-dimensional vectors are raised and lowered using metric coefficients $\delta_{\alpha\beta}=\delta^{\alpha\beta}$, i.e. there is no difference between covariant and contravariant components.

As mentioned previously, we do not assume the smallness of fluctuations for any type of perfect fluids. Therefore, in contrast to equation (2.9) of \cite{Claus1}, where fluctuations of the nonlinear perfect fluid are small quantities, herein we avoid the replacement of the combination  $(\varepsilon_J+ p_J)\tilde v_J^{\alpha}$ by $(\overline\varepsilon_J+ \overline p_J)\tilde v_J^{\alpha}$. Pressureless matter is taken in the form of discrete point-like masses with comoving mass density
\be{2.5}
\rho_M\equiv \sum_n m_n \delta({\bf r}-{\bf r}_n) \equiv \sum_n \rho_n\, ,
\ee
and its averaged energy density $\overline\varepsilon_M =\overline\rho_M c^2/a^3$. For such a component, the energy density fluctuation reads \cite{EZcosm1,EZcosm2,Chisari}
\be{2.6}
\delta\varepsilon_M=\frac{c^2}{a^3}\delta\rho_M+\frac{3\overline\rho_M c^2}{a^3}\Phi\, ,
\ee
with $\delta \rho_M\equiv \rho_M-\overline\rho_M$. This expression should be substituted into the right-hand side (RHS) of Eq.~\rf{2.3}. It is worth noting that we have dropped the term $\propto \delta\rho_M\Phi$. The point is that $\delta\rho_M$ is already a source for the metric correction $\Phi$. Therefore, in the perturbed Einstein equations, the product $\delta\rho_M\Phi$ results in corrections of the second order \cite{EinBril2}. As for the linear perfect fluid, the energy density can be considered in the form \cite{Claus1,Claus2,Eingorn2}
\ba{2.7}
\varepsilon_I&=&\frac{A_I}{a^{3(1+\omega_I)}}+3(1+\omega_I)\overline\varepsilon_I\Phi\,\nn\\
&=&\frac{\overline A_I}{a^{3(1+\omega_I)}}+\frac{\delta
	A_I}{a^{3(1+\omega_I)}}+\frac{3(1+\omega_I)\overline A_I}{a^{3(1+\omega_I)}}\Phi\, ,
\ea
where $A_I\equiv\ov A_I+\delta A_I$ and $\overline A_I =$ const. Since each matter component separately satisfies the energy conservation equation (see Eq. (A.20) in \cite{Claus1}) 
\ba{2.8}
&&\varepsilon'+3\mathcal{H}(\varepsilon+p)-3(\varepsilon+ p)\Phi'+\nabla\left[(\varepsilon+ p)\tilde{\mathbf{v}}\right]+\nabla\left[p\mathbf{B}\right]\,\nn\\
&&=0 \, ,\ea
in which $\varepsilon$ represents any of the individual components and $\Phi$ and $\mathbf B$ are the total potentials produced by the combination of components,
one can easily show that the function $A_I$ fulfills
\be{2.9}
A'_I+(1+\omega_I)\nabla\left(A_I\tilde{\bf v}_I\right)=0\, .
\ee
For the averaged quantities, Eq. \rf{2.8} yields
\be{2.10} \overline\varepsilon'+3\mathcal{H}(\overline\varepsilon+\overline p)=0\, ,\ee
and evidently, $\ov\varepsilon_I=\ov A_I/a^{3(1+\omega_I)}$ satisfies this equation. 

Let us now turn to the nonlinear perfect fluid with EoS $p_J=f_J(\varepsilon_J)$, where $f_J$ represents some nonlinear function. Since the fluctuations in the energy density and pressure are not restricted to small values, it is no longer possible to substitute $\ov p_J=f(\ov\varepsilon_J)$ for the background pressure; we need to proceed in a rather different way. Similar to Eqs. \rf{2.6} and \rf{2.7}, we consider the energy density in the form
\be{2.11}
\varepsilon_J=F_J+ 3\left(\varepsilon_J+ p_J\right)\Phi\, , 
\ee 
where $F_J$ is an unknown function for which we will derive an equation subsequently. Provided that $|\Phi|\ll1$, expanding the quantities $\varepsilon_J$ and $p_J$ accordingly, we may write
\ba{2.12} \varepsilon_J &=&F_J+3\left[F_J+f_J(F_J)\right]\Phi\, ,\\
\label{2.13}
p_J&=&\left.f_J(F_J)+3\frac{\partial f_J}{\partial\varepsilon_J}\right|_{\varepsilon_J=F_J}\left[F_J+f_J(F_J)\right]\Phi\, .
\ea
It is naturally demanded that the energy density \rf{2.11} satisfies the conservation equation \rf{2.8}. In this connection, we substitute \rf{2.12} and \rf{2.13} into \rf{2.8}, which yields
\ba{2.14}
&&\left.
\left[{F_J}'+3\mathcal{H}\left(F_J+f_J(F_J)\right)\right]\left[1+3\Phi\left(1+\frac{\partial f_J}{\partial\varepsilon_J}\right|_{\varepsilon_J=F_J}\right)\right] \nn\\
&+&\nabla\left[\left(F_J+f_J(F_J)\right)\tilde{\mathbf{v}}_J\right]+\mathbf{B}\nabla f_J(F_J)=0\, .
\ea
In obtaining the above expression we have neglected the terms quadratic with respect to $\Phi$ and used the relation $\nabla \mathbf{B}=0$ as well as $f'(F)=\partial f/\partial\varepsilon |_{\varepsilon =F} F'$.

Now, we decompose the functions $F_J$ and $f_J(F_J)$ into average values and fluctuations as
\be{2.15} 
F_J=\overline{F_J}+\delta F_J\, , \quad f_J(F_J)=\overline{f_J(F_J)}+\delta f_J\, ,
\ee
where both quantities $\overline{F_J}$ and $ \overline{f_J(F_J)}$ depend only on time. Obviously, $\ov\varepsilon_J=\overline{F_J}$ and $\ov p_J= \overline{f_J(F_J)}$ and hence the background equation \rf{2.10} for the ``$J$"-component reads
\be{2.16}
\overline{F_J}'+3\mathcal{H}\left(\overline{F_J}+\overline{f_J(F_J)}\right)=0 \, .
\ee
Substituting the decomposed functions \rf{2.15} into \rf{2.14} and taking into account \rf{2.16}, we get
\ba{2.17} 
&{}&
\left[{\delta F_J}'+3\mathcal{H}\left(\delta F_J+\delta f_J\right)\right]\,\nn\\
&\times&\left.\left[1+3\Phi\left(1+\frac{\partial f_J}{\partial\varepsilon_J}\right|_{\varepsilon_J=\overline F_J+\delta F_J}\right)\right]\,\nn\\
&+&\nabla\left[\left(F_J+f_J(F_J)\right)\tilde{\mathbf{v}}_J\right]+\mathbf{B}\nabla \delta f_J=0\, .
\ea

Before proceeding further, we pause to make a few important comments.
First, on large/cosmological scales the quantities $\Phi, B,\tilde v$
are of the same order of smallness $\epsilon$, i.e. $\Phi\sim B\sim \tilde v \sim \epsilon \ll1$. Meanwhile at small/astrophysical distances, we have 
$\Phi \sim \epsilon$ and $B\sim \tilde v \Phi$. Second, the perfect fluid is considered to
behave the ``normal" way. By this we mean that the squared speed of sound $c_s^2=\delta p/\delta \varepsilon \sim \partial f_J/\partial \varepsilon_J \lesssim 1$ and, additionally, the ratio of pressure fluctuations over pressure is of the order of its energy density counterpart:
$\delta\varepsilon/\varepsilon \sim \delta p/p$ $\;\Rightarrow \;$  $\delta\varepsilon_J/\varepsilon_J \sim \delta f_J/f_J$. Third, we exploit a useful estimate\footnote{\label{1}At small scales, $\delta \varepsilon/\varepsilon \sim 1$, thus the peculiar velocities comply with ${\tilde v}^2\sim \epsilon$. At large scales, as $\delta \varepsilon/\varepsilon \sim \epsilon$,  one gets $\tilde v \sim \epsilon$ instead.} \cite{Baumann,EinBril2}
\be{2.18} 
\Phi\frac{\delta\varepsilon}{\varepsilon}\sim {\tilde v}^2\, .
\ee
Keeping these in mind, the terms multiplied by the scalar perturbation $\Phi$ can be safely neglected in comparison to the rest of the expression in \rf{2.17}. For $\delta f_J B\ll f_J \tilde{v}_J $, we find that the left-hand side (LHS) is further simplified to yield  
\be{2.19}
{\delta F_J}'+3\mathcal{H}\left(\delta F_J+\delta f_J\right)+\nabla\left[\left(F_J+f_J(F_J)\right)\tilde{\mathbf{v}}_J\right]=0\, ,\ee
which, combined with \rf{2.16}, represents the sought-for equation for the function $F_J$.

Now, we return to Eqs. \rf{2.3} and \rf{2.4} for the potentials $\Phi$ and $\mathbf B$. Substituting \rf{2.7} together with the definitions from \rf{2.12} and \rf{2.13} into Eq. \rf{2.4}, we find
\ba{2.20}
&&\frac{1}{4}\triangle B_\alpha+\frac{\partial}{\partial x^\alpha}\left(\Phi'+\mathcal{H}\Phi\right)\,\nn\\
&=&\frac{1}{2}\kappa a^2\left(-\frac{c^2}{a^3}\sum_n\rho_n\tilde{v}^\alpha_n+\frac{\overline\rho_Mc^2}{a^3}B_\alpha\right. \,\nn\\
&-&\sum_I\frac{1+\omega_I}{a^{3\left(1+\omega_I\right)}}A_I\tilde{v}^\alpha_I+\sum_I\left(\overline\varepsilon_I+\overline p_I\right)B_\alpha\,\nn\\
&-&\left.\sum_J\left(F_J+f_J(F_J)\right)\tilde{v}^\alpha_J+\sum_J\left(\overline\varepsilon_J+\overline p_J\right)B_\alpha\right)\, ,\nn\\
\ea
where we have taken into account that the terms proportional to the products $\Phi\tilde{v}^\alpha_{I,J}$ are to be neglected in the first order. Moving further, we decompose the terms with $\tilde{v}^\alpha_{n,I,J}$ on the RHS into their longitudinal and transverse parts as (see Eqs. (2.24) and (2.25) in \cite{Claus1})
\ba{2.21}
\sum_n\rho_n\tilde {\bf v}_n &&= \nabla \Xi+\left(\sum_n\rho_n\tilde {\bf v}_n-\nabla\Xi\right)\, ,\,
 \nabla \left(\sum_n\rho_n\tilde {\bf v}_n\right)=\triangle\Xi\, ,\nn\\
\\
\label{2.22}A_I\tilde {\bf v}_I&&=\nabla\xi_I+\left(A_I\tilde {\bf v}_I-\nabla\xi_I\right)\, ,
\, \nabla(A_I\tilde {\bf v}_I)=\triangle\xi_I\, ,\ea
\ba{2.23} &&\left(F_J+f_J(F_J)\right)\tilde{\mathbf{v}}_J=\nabla\zeta_J+\left[\left(F_J+f_J(F_J)\right)\tilde{\mathbf{v}}_J-\nabla\zeta_J\right]\, ,\nn\\
 &&\qquad\qquad\nabla \left[\left(F_J+f_J(F_J)\right)\tilde{\mathbf{v}}_J\right]=\triangle\zeta_J.
\ea
Here the function $\Xi$ has the form \cite{Eingorn1}
\be{2.24}
\Xi=\frac{1}{4\pi}\sum\limits_nm_n\frac{({\bf r}-{\bf r}_n)\tilde {\bf v}_n}{|{\bf r}-{\bf r}_n|^3}\, ,\ee
whereas $\xi_I$ and $\zeta_J$ are to be determined numerically.
Using this system of equations, we  split Eq.~\rf{2.20} into scalar and vector parts so as to obtain the equations
\be{2.25}
\Phi'+\mathcal{H}\Phi=-\frac{\kappa c^2}{2a}\Xi-\frac{\kappa}{2}\sum_I\frac{1+\omega_I}{a^{1+3\omega_I}}\xi_I-\frac{\kappa a^2}{2}\sum_J\zeta_J \, ,
\ee
\ba{2.26}
&&\frac{1}{4}\triangle\mathbf{B}\,\nn\\
&-&\left[\frac{\kappa\overline\rho_M c^2}{2a} +\frac{\kappa a^2}{2}\sum_I\left(\overline\varepsilon_I+\overline p_I\right) +\frac{\kappa a^2}{2}\sum_J\left(\overline\varepsilon_J+\overline p_J\right) \right]\mathbf{B}\nn\\
&=&-\frac{\kappa c^2}{2a}\left(\sum_n\rho_n\tilde{\mathbf{v}}_n-\nabla\Xi\right)-\frac{\kappa}{2}\sum_I\frac{1+\omega_I}{a^{1+3\omega_I}}\left(A_I\tilde{\mathbf{v}}_I-\nabla\xi_I\right)\,\nn\\
&-&\frac{\kappa a^2}{2}\sum_J\left[\left(F_J+f_J(F_J)\right)\tilde{\mathbf{v}}_J-\nabla\zeta_J\right]\, . 
\ea
Substituting \rf{2.25} into Eq. \rf{2.3} and taking into account Eqs. \rf{2.6}, \rf{2.7} together with the  expression
\be{2.27}
\delta\varepsilon_J\equiv\varepsilon_J-\ov\varepsilon_J=\delta F_J+3\left(\overline F_J+\overline{f_J(F_J)}\right)\Phi\, ,
\ee
we get the equation for the potential $\Phi$:
\ba{2.28}
&&\triangle\Phi\,\nn\\
&-&\frac{3}{2}\kappa a^2 \left[\frac{\overline\rho_Mc^2}{a^3}+\sum_I\frac{(1+\omega_I)\overline{A}_I}{a^{3(1+\omega_I)}} +
\sum_J\left(\overline F_J+\overline{f_J(F_J)}\right)\right]\Phi\nn\\
&=&\frac{1}{2}\kappa a^2 \left[\frac{c^2}{a^3}\delta\rho_M+\sum_I\frac{\delta A_I}{a^{3(1+\omega_I)}}
+\sum_J\delta F_J\right]\nn\\
&-&\frac{3\kappa c^2\mathcal{H}}{2a}\Xi-\frac{3\mathcal{H}\kappa}{2}\sum_I\frac{1+\omega_I}{a^{1+3\omega_I}}\xi_I-\frac{3\mathcal{H}\kappa a^2}{2}\sum_J\zeta_J\, .
\ea

It is possible to reformulate \rf{2.26} and \rf{2.28} so that we have
\ba{2.29}
&{}&\triangle\Phi-\frac{a^2}{\lambda^2}\Phi\,\nn\\
&=&\frac{\kappa c^2}{2a}\delta\rho_M+\frac{\kappa a^2}{2}\sum_I\frac{\delta A_I}{a^{3(1+\omega_I)}}+\frac{\kappa a^2}{2}\sum_J\delta F_J\,\nn\\
&-&\frac{3\kappa c^2\mathcal{H}}{2a}\Xi-\frac{3\mathcal{H}\kappa}{2}\sum_I\frac{1+\omega_I}{a^{1+3\omega_I}}\xi_I-\frac{3\mathcal{H}\kappa a^2}{2}\sum_J\zeta_J\, 
\ea
and
\ba{2.30}
&{}&
\frac{1}{4}\triangle\mathbf{B}-\frac{a^2}{3\lambda^2}\mathbf{B}\,\nn\\
&=&-\frac{\kappa c^2}{2a}\left(\sum_n\rho_n\tilde{\mathbf{v}}_n-\nabla\Xi\right)-\frac{\kappa}{2}\sum_I\frac{1+\omega_I}{a^{1+3\omega_I}}\left(A_I\tilde{\mathbf{v}}_I-\nabla\xi_I\right)\,\nn\\
&-&\frac{\kappa a^2}{2}\sum_J\left[\left(F_J+f_J(F_J)\right)\tilde{\mathbf{v}}_J-\nabla\zeta_J\right] \, ,
\ea
where
\ba{2.31}
\lambda&\equiv&\left[\frac{3\kappa\overline\rho
	_Mc^2}{2a^3}\right.\,\nn\\
	&+&\left.\frac{3\kappa}{2}\sum_I\frac{(1+\omega_I)\overline{A}_I}{a^{3(1+\omega_I)}}+\frac{3\kappa}{2}\sum_J\left(\overline F_J+\overline{f_J(F_J)}\right)\right]^{-1/2}\, .\ea
With only a single ``$J$"-component\footnote{\label{2}As we have noted above, the ``$M$"- and ``$I$"-components can be considered as particular cases of the ``$J$"-component. For example, for the ``$I$"-component with the EoS $p_I=\omega_I\varepsilon_I$, we have $\partial f_I/\partial\varepsilon_I=\omega_I, F_I\equiv A_I/a^{3(1+\omega_I)}$ and $A_I$ satisfies Eq. \rf{2.9}. Hence, in this particular case Eq. \rf{2.19} is reduced to \rf{2.9}.} in the universe, Eqs. \rf{2.29} and \rf{2.30} are reduced to 
\be{2.32}
\triangle\Phi-\frac{a^2}{\lambda^2}\Phi=\frac{\kappa a^2}{2}\delta F
_J-\frac{3\mathcal{H}\kappa a^2}{2}\zeta_J\, \ee
and
\be{2.33}
\frac{1}{4}\triangle\mathbf{B}-\frac{a^2}{3\lambda^2}\mathbf{B}=-\frac{\kappa a^2}{2}\left[\left(F_J+f_J(F_J)\right)\tilde{\mathbf{v}}_J-\nabla\zeta_J\right] \, ,
\ee
respectively, accompanied by the evident redefinition of  $\lambda$. It is remarkable that in these equations the scalar and vector perturbations are decoupled. Obviously, we get the same Eq.~\rf{2.29} if we discard the vector perturbations and work from the very beginning in the conformal Newtonian gauge. It is well known that in this gauge the perturbations coincide with the Bardeen gauge-invariant ones.

We also need the momentum conservation equation for the ``$J$"-component. Using (A.23) from \cite{Claus1} and eliminating the terms proportional to $\Phi^2, \Phi B, \Phi\tilde v, B\tilde v$, which are of the higher order of smallness, we obtain
\ba{2.34}
&&(F_J+f_J(F_J))\mathbf{B}'\,\nn\\
&+&\left.\left[\left(1+\frac{\partial f_J}{\partial\varepsilon_J}\right|_{\varepsilon_J=F_J}\right)F'_J+4\mathcal{H}(F_J+f_J(F_J))\right]\mathbf{B} \,\nn\\
&-&\left[(F_J+f_J(F_J))\tilde{\mathbf{v}}_J-\nabla\zeta_J\right]'-4\mathcal{H}\left[(F_J+f_J(F_J))\tilde{\mathbf{v}}_J-\nabla\zeta_J\right]\,\nn\\
&-&\left.\nabla\left(\zeta_J'+4\mathcal{H}\zeta_J+f_J(F_J)+3\frac{\partial f_J}{\partial\varepsilon_J}\right|_{\varepsilon_J=F_J}\left[F_J+f_J(F_J)\right]\Phi\right)\,\nn\\
&-&(F_J+f_J(F_J))\nabla\Phi=0 \, .
\ea

This equation may readily be used to define the time derivative of peculiar velocity and to this end, one may replace $\mathbf{B}'$ with the help of the equation $\mathbf{B}'=-2\mathcal{H}\mathbf{B}$ \cite{Eingorn1,Claus1}.

To conclude this section, we briefly consider  the Chaplygin gas model \cite{Chap1,Chap2,Chap3} as a particular example of the ``$J$"-type nonlinear perfect fluid. The EoS of the modified generalized Chaplygin gas has the form \cite{Chap4}
\be{2.35}
p=f(\varepsilon)=\beta\varepsilon-\left(1+\beta\right)\frac{A}{\varepsilon^\alpha} \, , \quad \beta ,A, \alpha :\mathrm{const}\, ,
\ee
for which the background quantities read
\be{2.36}
\overline\varepsilon=\overline F \, ,\quad \overline p=\overline{f(F)}=\beta\overline F-A\left(1+\beta\right)\overline{\left(\frac{1}{F^{\alpha}}\right)} \, . 
\ee
These satisfy the conservation equation \rf{2.16}
\ba{2.37}
\overline F'+3\mathcal{H}\left[\left(1+\beta\right)\overline F-A\left(1+\beta\right)\overline{\left(\frac{1}{F^{\alpha}}\right)}\right]=0\, , 
\ea
which serves to define $\ov F$.
Unfortunately, it cannot be solved analytically since averaging of the unknown function $1/F^{\alpha}$ is not possible. Therefore, in the case of perfect fluids with nonlinear EoS, models of interest should be investigated numerically. The general strategy is as follows: first, one defines the initial values for $F, \tilde v$ and the scale factor $a$. Then, performing iteration, one solves the system of 6 equations which are the energy conservation for the background and perturbed values \rf{2.16} and \rf{2.19}, the momentum conservation equation \rf{2.34}, the Friedmann equation \rf{2.1} and equations for the potentials $\Phi$ and $\mathbf B$ obtained in \rf{2.32} and \rf{2.33}, respectively. The explicit algorithm for the numerical simulation consists in the following: one starts with solving Eqs.~\rf{2.32} and \rf{2.33} for $\Phi$ and $\mathbf B$ at some moment of time for known sources of these potentials, and then, with the help of their values, determines $F_J$ and $\tilde{\mathbf{v}}_J$ as well as the scale factor $a$ at the next moment from \rf{2.16}, \rf{2.19} and \rf{2.34}, as well as \rf{2.1}. After such a straightforward procedure, everything is ready for the new iteration step.

\section{Conclusion}

In our paper we have studied a universe consisting of three types of components. It contains dust-like matter (denoted by the index ``$M$") in the form of discrete inhomogeneities (e.g., galaxies, galaxy groups and clusters), perfect fluids with linear EoS $p_I=\omega_I\varepsilon_I\, \
(\omega_I=\mathrm{const})$ and perfect fluids with arbitrary nonlinear EoS $p_J=f_J(\varepsilon_J)$. The
background spacetime geometry is defined by the Friedmann-Lema\^{\i}tre-Robertson-Walker metric. All three components have been considered to have arbitrary energy density contrasts. Within the cosmic screening approach, we have developed the theory of scalar and vector perturbations and obtained a system of equations that enables the numerical simulation of models with an arbitrary form of the function $f(\varepsilon)$. Since we have not assumed the smallness of energy density fluctuations, these equations are valid on both small/astrophysical and large/cosmological scales. 

Additionally, we have checked some of the important auxiliary equations to demonstrate that they are indeed satisfied (up to the adopted accuracy) for arbitrary density contrasts.
\appendix

\section{Check of the equation \rf{2.25}}\label{app}
\renewcommand{\theequation}{A.\arabic{equation}}
The cosmic screening approach \cite{Eingorn1,Claus1} applies to all scales down to astrophysical distances where energy density fluctuations of the perfect fluids are no longer small quantities. In this appendix we intend to demonstrate that Eq.~\rf{2.25} holds for arbitrary values of density contrasts. For this purpose, we will be preserving the terms of the type $\delta\rho\Phi, \delta A\Phi, \delta F\Phi$ which can exceed $\ov\rho\Phi, \ov A\Phi, \ov F\Phi$ at small scales. 

We have already noted that ``$M$"- and ``$I$"-components can be considered as particular cases of the ``$J$"-component. Hence, it would indeed be sufficient to limit ourselves to the ``$J$"-components in \rf{2.25}. Nevertheless, our reasoning will be clearer if we first prove the equation for the ``$M$"-component, then for the ``$I$"-component, and only after that for the ``$J$"-component. Considering the ``M"-component separately from the other two is especially justified by the fact that there exist exact expressions for the potentials $\Phi$ and $\mathbf B$ in the corresponding model.

\subsection{Pressureless matter}

We consider the ``$M$"-component in the form of discrete point-like masses with the comoving mass density \rf{2.5}. For this case, the potential $\Phi$ in Fourier space reads \cite{Eingorn1}

\ba{a.1} 
\widehat{\Phi}&=&-\frac{\kappa c^2}{2a}\left(k^2+\frac{3\kappa \overline{\rho}c^2}{2a}\right)^{-1}\,\nn\\
&\times&\left[\widehat{\delta\rho} +
3i\mathcal{H}\sum_{n}m_{n}e^{-i{\bf k}{\bf r}_{n}}\frac{({\bf k}\tilde{{\bf v}}_{n})}{k^2} \right]\, ,
\ea
where
\be{a.2} 
\widehat{\delta\rho} = \sum_{n}\widehat{\rho}_n - \widehat{\ov{\rho}}=\sum_{n}m_{n}e^{-i{\bf k}{\bf r}_{n}} -\overline{\rho}(2\pi)^3 \delta({\bf k})\, .
\ee
Above and hereafter the hat denotes the Fourier-transforms.
In the current arrangement, Eq. \rf{2.25} takes the form
\be{a.3} 
\widehat{\Phi}'+\mathcal{H}\widehat{\Phi}=-\frac{\kappa c^2}{2a}\widehat{\Xi} \, ,
\ee
where \cite{Eingorn1}
\be{a.4}
\widehat{\Xi} =
-\frac{i}{k^2}\sum_{n}m_{n}({\bf k}\tilde{{\bf v}}_{n})e^{-i{\bf k}{\bf r}_{n}}=-\frac{i}{k^2}\sum_{n}\widehat{\rho}_n ({\bf k}\tilde{{\bf v}}_{n})\, .
\ee
Differentiating the expression \rf{a.1} with respect to conformal time and  neglecting the terms quadratic in $\tilde {\bf v}_n$, we obtain
\ba{a.5} 
\widehat{\Phi}' &=& \frac{\kappa c^2}{2a}\mathcal{H}\,\nn\\
&\times&\left(k^2+\frac{3\kappa \overline{\rho}c^2}{2a}\right)^{-1}\left[\widehat{\delta\rho} +
3i\frac{\mathcal{H}}{a}\sum_{n}m_{n}e^{-i{\bf k}{\bf r}_{n}}\frac{({\bf k}a\tilde{{\bf v}}_{n})}{k^2} \right] \nn \\
&-& \frac{3\kappa^2\overline{\rho} c^4}{4a^2}\mathcal{H}\,\nn\\
&\times&\left(k^2+\frac{3\kappa\overline{\rho} c^2}{2a}\right)^{-2}\left[\widehat{\delta\rho} +
3i\frac{\mathcal{H}}{a}\sum_{n}m_{n}e^{-i{\bf k}{\bf r}_{n}}\frac{({\bf k}a\tilde{{\bf v}}_{n})}{k^2} \right] \nn \\
&-& \frac{\kappa c^2}{2a}\left(k^2+\frac{3\kappa \overline{\rho}c^2}{2a}\right)^{-1}\,\nn\\
&\times&\left[\left(\widehat{\delta\rho}\right)' +
3i\left(\frac{\mathcal{H}}{a}\right)'\sum_{n}m_{n}e^{-i{\bf k}{\bf r}_{n}}\frac{({\bf k}a\tilde{{\bf v}}_{n})}{k^2} \right. \nn \\
&+& \left. 3i\frac{\mathcal{H}}{a}\sum_{n}m_{n}e^{-i{\bf k}{\bf r}_{n}}\frac{{\bf k}(a\tilde{{\bf v}}_{n})'}{k^2}\right]\, .
\ea
Now, taking into account the relation \cite{Eingorn1}
\ba{a.6} 
\sum\limits_n\widehat{\rho}_n(a\tilde{{\bf v}}_{n})' &=& \sum_{n}m_{n}e^{-i{\bf k}{\bf r}_{n}} (a\tilde{{\bf v}}_{n})'\,\nn\\
&=& -a\left[\rho \nabla\Phi\right]\widehat{} +\left[\rho (a{\bf B})'\right]\widehat{}\, ,\ea
and expressing the time derivative $\left(\widehat{\delta\rho}\right)'$ using \rf{a.2}, we reformulate \rf{a.5} as
\ba{a.7} 
\widehat{\Phi}' &=& -\frac{\kappa c^2}{2a}\left(k^2+\frac{3\kappa \overline{\rho}c^2}{2a}\right)^{-1}\,\nn\\
&\times&\left[-\mathcal{H}\widehat{\delta\rho} -i{\bf k}
\sum\limits_n\hat{\rho}_n\tilde{{\bf v}}_{n} -3i\frac{\mathcal{H}^2}{a}\sum_{n}\widehat{\rho}_{n}\frac{({\bf k}a\tilde{{\bf v}}_{n})}{k^2} \right. \nn \\
&+&  3i\left(\frac{\mathcal{H}}{a}\right)'\sum_{n}\widehat{\rho}_{n}\frac{({\bf k}a\tilde{{\bf v}}_{n})}{k^2} +3\mathcal{H}\overline{\rho}\widehat{\Phi}\,\nn\\
&-&\left.3i\frac{\mathcal{H}}{k^2}{\bf k}\left[\delta\rho \nabla\Phi\right]\hat{} +3i\frac{\mathcal{H}}{ak^2}{\bf k}\left[\delta\rho (a{\bf B})'\right]\widehat{}\right]\nn \\
&-& \frac{3\kappa^2\overline{\rho} c^4}{4a^2}\mathcal{H}\,\nn\\
&\times&\left(k^2+\frac{3\kappa\overline{\rho} c^2}{2a}\right)^{-2}\left[\widehat{\delta\rho} +
3i\frac{\mathcal{H}}{a}\sum_{n}\hat{\rho}_{n}\frac{({\bf k}a\tilde{{\bf v}}_{n})}{k^2} \right]\, .
\ea

The term $-3i\cfrac{\mathcal{H}}{k^2}{\bf k}\left[\delta\rho \nabla\Phi\right]\widehat{}\sim\mathcal{H}\delta\rho\cdot \epsilon$ in the square brackets can be
neglected in comparison with the term $-\mathcal{H}\widehat{\delta\rho}\sim\mathcal{H}\delta\rho$ (their ratio gives precisely the order of smallness $\epsilon$).
Similarly, the term $3i\cfrac{\mathcal{H}}{ak^2}{\bf k}\left[\delta\rho (a{\bf B})'\right]\widehat{}$ is much less than the term $-3i\cfrac{\mathcal{H}^2}{a}\sum\limits_{n}\widehat{\rho}_{n}\cfrac{({\bf k}a\tilde{{\bf v}}_{n})}{k^2}$. Their ratio is of the order of $\delta \rho B/\left(\rho
\tilde v\right)\sim B\tilde v/\left(\rho \tilde v^2/\delta\rho\right)\sim B\tilde v/\Phi \sim \epsilon$. Here, we use the relation  ${\bf B}'=-2\mathcal{H}{\bf B}$, the estimate \rf{2.18} and take into account that at small/astrophysical scales $B \sim \tilde v \Phi$ while at large/cosmological scales $B\sim \Phi$ (see also footnote \ref{1}). 

Substituting the expressions for $\widehat{\Phi}$ and $\widehat{\Phi}'$ into the left-hand side (LHS) of \rf{a.3}, we get
\ba{a.8} 
&{}&\widehat{\Phi}'+\mathcal{H}\widehat{\Phi} 
\,\nn\\
&=&-\frac{\kappa c^2}{2a}\left(k^2+\frac{3\kappa \overline{\rho}c^2}{2a}\right)^{-1}\,\nn\\
&\times&\left[-i\sum\limits_n\widehat{\rho}_n({\bf k}\tilde{{\bf v}}_{n})+
3i\left(\frac{\mathcal{H}}{a}\right)'\sum_{n}\widehat{\rho}_{n}\frac{({\bf k}a\tilde{{\bf v}}_{n})}{k^2} \right] \nn \\
&+& \frac{3\kappa^2\overline{\rho} c^4}{4a^2}\mathcal{H}\,\nn\\
&\times&\left(k^2+\frac{3\kappa \overline{\rho}c^2}{2a}\right)^{-2}\left[\widehat{\delta\rho} +
3i\mathcal{H}\sum_{n}m_{n}e^{-i{\bf k}{\bf r}_{n}}\frac{({\bf k}\tilde{{\bf v}}_{n})}{k^2}\right] \nn \\
&-& \frac{3\kappa^2\overline{\rho} c^4}{4a^2}\mathcal{H}\,\nn\\
&\times&\left(k^2+\frac{3\kappa\overline{\rho} c^2}{2a}\right)^{-2}\left[\widehat{\delta\rho} +
3i\frac{\mathcal{H}}{a}\sum_{n}\widehat{\rho}_{n}\frac{({\bf k}a\tilde{{\bf v}}_{n})}{k^2} \right] \nn \\
&=&
\frac{\kappa c^2}{2a^2}i\left(k^2+\frac{3\kappa \overline{\rho}c^2}{2a}\right)^{-1}\left[k^2-3a\left(\frac{\mathcal{H}}{a}\right)'\right]
\sum_{n}\widehat{\rho}_{n}\frac{({\bf k}a\tilde{{\bf v}}_{n})}{k^2} \nn \\
&=&
\frac{\kappa c^2}{2a^2}i\left(k^2+\frac{3\kappa \overline{\rho}c^2}{2a}\right)^{-1}\,\nn\\
&\times&\left[k^2+3\left(\mathcal{H}^2 -\mathcal{H}'\right)\right]
\sum_{n}\widehat{\rho}_{n}\frac{({\bf k}a\tilde{{\bf v}}_{n})}{k^2} \nn \\
&=& \frac{\kappa c^2}{2a^2}i\left(k^2+\frac{3\kappa \overline{\rho}c^2}{2a}\right)^{-1}\left(k^2+\frac{3\kappa \overline{\rho}c^2}{2a}\right)
\sum_{n}\widehat{\rho}_{n}\frac{({\bf k}a\tilde{{\bf v}}_{n})}{k^2} \,\nn\\
&=& \frac{\kappa c^2}{2a}i\sum_{n}\widehat{\rho}_{n}\frac{({\bf k}\tilde{{\bf v}}_{n})}{k^2}\, ,\ea
and this exactly coincides with the RHS of \rf{a.3} (see Eq.~\rf{a.4}).

\subsection{Perfect fluid with linear equation of state}

In this section we consider perfect fluids with linear EoS $p_I=\omega_I \varepsilon_I$,  $\omega_I=\mathrm{const}$. From Eq.~\rf{2.25} we obtain
\ba{a.9} 
\triangle\Phi' +\mathcal{H}\triangle\Phi = -\frac{1}{2}\kappa a^2\sum\limits_I \left(1+\omega_I\right)\triangle\zeta_I\, ,\ea
where
\be{a.10}
\zeta_I\equiv \frac{\xi_I}{a^{3(1+\omega_I)}}\, .
\ee
Our intention now is to prove that Eq. \rf{a.9} holds for arbitrary values of energy density fluctuations.

From Eq. \rf{2.29} and taking into account the definition \rf{2.31} for $\lambda$, we get
\ba{a.11} 
\triangle\Phi &=& \frac{3}{2}\kappa a^2 \left[\sum\limits_I \frac{1+\omega_I}{a^{3\left(1+\omega_I\right)}}\overline{A}_I\right]\Phi \,\nn\\
&+&\frac{1}{2}\kappa
a^2\sum\limits_I \cfrac{\delta A_I}{a^{3\left(1+\omega_I\right)}} - \frac{3}{2}\kappa a^2\mathcal{H}\sum\limits_I\left(1+\omega_I\right)\zeta_I\, .
\ea
The time derivative of this equation is
\ba{a.12} 
\triangle\Phi' 
&=& \frac{3}{2}\kappa a^2\left[\sum\limits_I \cfrac{1+\omega_I}{a^{3\left(1+\omega_I\right)}}\overline{A}_I\right]\left( \Phi' +2\mathcal{H}\Phi\right)\,\nn\\
&-&
\frac{9}{2}\kappa a^2\mathcal{H} \left[\sum\limits_I \cfrac{\left(1+\omega_I\right)^2}{a^{3\left(1+\omega_I\right)}}\overline{A}_I\right]\Phi\nn \\
&-& \frac{1}{2}\kappa a^2\mathcal{H}\sum\limits_I \cfrac{\delta A_I}{a^{3\left(1+\omega_I\right)}} + \frac{1}{2}\kappa a^2\sum\limits_I \cfrac{\delta
	A'_I}{a^{3\left(1+\omega_I\right)}}\,\nn\\
	&-&
\frac{3}{2}\kappa a^2\mathcal{H} \sum\limits_I \omega_I\cfrac{ \delta A_I}{a^{3\left(1+\omega_I\right)}}\nn \\
&-& \frac{3}{2}\kappa a^2\left(2\mathcal{H}^2+\mathcal{H}'\right)\sum\limits_I\left(1+\omega_I\right)\zeta_I\,\nn\\
&-&\frac{3}{2}\kappa
a^2\mathcal{H}\sum\limits_I\left(1+\omega_I\right)\zeta'_I\, ,
\ea
and substitution of \rf{a.11} and \rf{a.12} into the LHS of \rf{a.9} results in
\ba{a.13} &{}& \frac{3}{2}\kappa a^2\left[\sum\limits_I \cfrac{1+\omega_I}{a^{3\left(1+\omega_I\right)}}\overline{A}_I\right]\left( \Phi' +3\mathcal{H}\Phi\right)\,\nn\\
&-&
\frac{9}{2}\kappa a^2\mathcal{H} \left[\sum\limits_I \cfrac{\left(1+\omega_I\right)^2}{a^{3\left(1+\omega_I\right)}}\overline{A}_I\right]\Phi +
\frac{1}{2}\kappa a^2\sum\limits_I \cfrac{\delta A'_I}{a^{3\left(1+\omega_I\right)}} \nn \\
&-& \frac{3}{2}\kappa a^2\mathcal{H} \sum\limits_I \omega_I\cfrac{ \delta A_I}{a^{3\left(1+\omega_I\right)}}\,\nn\\
&-&\frac{3}{2}\kappa a^2\left(3\mathcal{H}^2+
\mathcal{H}'\right)\sum\limits_I\left(1+\omega_I\right)\zeta_I\,\nn\\
&-&\frac{3}{2}\kappa a^2\mathcal{H}\sum\limits_I\left(1+\omega_I\right)\zeta'_I \nn \\
&=& -\frac{1}{2}\kappa a^2\sum\limits_I \left(1+\omega_I\right)\triangle\zeta_I \, ,\ea
where the terms $\propto \kappa a^2\mathcal{H} \delta A_I/a^{3\left(1+\omega_I\right)}$ have been cancelled.

In order to proceed, we need to determine the quantities $\delta A'_I$ and $\zeta'_I$. For this purpose, we employ \rf{2.7} as the energy density $\varepsilon_I = \overline{\varepsilon}_I +\delta\varepsilon_I$ of the ``$I$"-component and re-express Eq.~\rf{2.8} in terms of this component only, which reads 
\ba{a.14} 
&&\cfrac{\delta A'_I}{a^{3\left(1+\omega_I\right)}} -3\left(1+\omega_I\right)\cfrac{\delta A_I}{a^{3\left(1+\omega_I\right)}}\Phi'
+\left(1+\omega_I\right)\triangle\zeta_I\,\nn\\
&&+\nabla\left[\delta p_I {\bf B}\right] =0 \, .\ea
Here we have taken into account the definition \rf{2.22} as well as the transverse gauge condition $\nabla\mathbf{B}=0$. It is worth noting that neither of the cases $|\omega_I|\gg 1$ or $|1+\omega_I|\ll 1$ are being considered in the current configuration. Now, the second term in \rf{a.14} can be neglected in comparison with the first one, since $\delta A_I\Phi'=\left(\delta A_I\Phi\right)'-\delta A'_I\Phi\ll \delta A'_I$: 
\be{a.15} 
\cfrac{\delta A'_I}{a^{3\left(1+\omega_I\right)}} = -\left(1+\omega_I\right)\triangle\zeta_I
-\cfrac{\omega_I}{a^{3\left(1+\omega_I\right)}}\nabla\left[\delta A_I{\bf B}\right] \, .
\ee
From $\triangle\zeta_I=\nabla\left(A_I\tilde {\bf v}_I/a^{3(1+\omega_I)}\right)$, it is possible to further conclude that the second term on the RHS may again be neglected in comparison with the first one, if we also take into account the inequality $(\delta
A_I/A_I)B\ll \tilde v_I$ (similar to the ones used in previous sections).  Consequently,
\be{a.16} 
\cfrac{\delta A'_I}{a^{3\left(1+\omega_I\right)}} = -\left(1+\omega_I\right)\triangle\zeta_I \, .
\ee

Further on, in order to find $\zeta'_I$, we use the momentum conservation equation (A.23) in \cite{Claus1} applied to the ``$I$"-component (which is the analogue of \rf{2.34} for the ``$J$"-component):
\ba{a.17} 
&{}&\left(1+\omega_I\right)\left[\varepsilon_I{\bf B}\right]' -\left(1+\omega_I\right)\left[\varepsilon_I\tilde{\bf v}_I\right]'
+4\mathcal{H}\left(1+\omega_I\right)\varepsilon_I{\bf B} \nn\\ 
&-&4\mathcal{H}\left(1+\omega_I\right)\varepsilon_I\tilde{\bf v}_I-\omega_I\nabla\varepsilon_I
-\left(1+\omega_I\right)\varepsilon_I\nabla\Phi =0\, .
\nn\\\ea
Using the definition \rf{2.7}, eliminating the second order terms, replacing ${\bf B}'$ with the expression $-2\mathcal{H}{\bf B}$ and then acting by $\nabla$, we obtain
\ba{a.18} 
&{}& \cfrac{\left(1+\omega_I\right)}{a^{3\left(1+\omega_I\right)}}\nabla\left[\delta A'_I{\bf B}\right] -
\cfrac{3\left(1+\omega_I\right)^2}{a^{3\left(1+\omega_I\right)}}\mathcal{H}\nabla\left[\delta A_I{\bf B}\right]  \nn \\
&+&2\mathcal{H}\cfrac{\left(1+\omega_I\right)}{a^{3\left(1+\omega_I\right)}}\nabla\left[\delta A_I{\bf B}\right] \,\nn\\
&-&\left(1+\omega_I\right)\triangle\zeta_I'-4\mathcal{H}\left(1+\omega_I\right)\triangle\zeta_I \,\nn\\
&-&\cfrac{\omega_I}{a^{3\left(1+\omega_I\right)}}\triangle\delta
A_I -
\cfrac{3\omega_I\left(1+\omega_I\right)\overline{A}_I}{a^{3\left(1+\omega_I\right)}}\triangle\Phi\nn \\
&-& \cfrac{\left(1+\omega_I\right)\overline{A}_I}{a^{3\left(1+\omega_I\right)}}\triangle\Phi
-\cfrac{\left(1+\omega_I\right)}{a^{3\left(1+\omega_I\right)}}\nabla\left[\delta A_I\nabla\Phi\right] =0\, .\ea
Since  $\triangle\zeta_I=\nabla\left(A_I\tilde {\bf v}_I/a^{3(1+\omega_I)}\right)$,  all terms in the first line can be neglected in comparison with the first and second terms in the second line (the ratios reduce to \mbox{$\sim(\delta
A_I/A_I)(B/\tilde v_I)\sim B\tilde v_I/\Phi$}). Hence,
\ba{a.19} 
&&-\left(1+\omega_I\right)\triangle\zeta_I' -4\mathcal{H}\left(1+\omega_I\right)\triangle\zeta_I
-\cfrac{\omega_I}{a^{3\left(1+\omega_I\right)}}\triangle\delta A_I\nn\\
&-&\cfrac{\left(1+3\omega_I\right)\left(1+\omega_I\right)\overline{A}_I}{a^{3\left(1+\omega_I\right)}}\triangle\Phi
-\cfrac{\left(1+\omega_I\right)}{a^{3\left(1+\omega_I\right)}}\nabla\left[\delta A_I\nabla\Phi\right] =0\, ,\nn\\
\ea
and we arrive at 
\ba{a.20} 
\left(1+\omega_I\right)\zeta_I' &=& -4\mathcal{H}\left(1+\omega_I\right)\zeta_I -\cfrac{\omega_I}{a^{3\left(1+\omega_I\right)}}\delta A_I\,\nn\\
&-&\cfrac{\left(1+3\omega_I\right)\left(1+\omega_I\right)\overline{A}_I}{a^{3\left(1+\omega_I\right)}}\Phi\,\nn\\
&-&\cfrac{\left(1+\omega_I\right)}{a^{3\left(1+\omega_I\right)}}\triangle^{-1}\nabla\left[\delta A_I\nabla\Phi\right]\, .\nn\\
\ea

Substituting Eqs. \rf{2.25} (for the  ``$I$"-component), \rf{a.16} and \rf{a.20} (for $\Phi'$, $\delta A'_I$, and $\zeta'_I$, respectively) into Eq.~\rf{a.13} and simplifying, which amounts to cancelling the terms \mbox{$\propto \kappa a^2\mathcal{H}
\sum\limits_I \omega_I\delta A_I/a^{3\left(1+\omega_I\right)}$}, we get
\ba{a.21} 
&{}& \frac{3}{2}\kappa a^2\left[\sum\limits_I \cfrac{1+\omega_I}{a^{3\left(1+\omega_I\right)}}\overline{A}_I\right]\left(2\mathcal{H}\Phi -
\frac{1}{2}\kappa a^2\sum\limits_I \left(1+\omega_I\right)\zeta_I\right)\,\nn\\
&-&\frac{9}{2}\kappa a^2\mathcal{H}
\left[\sum\limits_I \cfrac{\left(1+\omega_I\right)^2}{a^{3\left(1+\omega_I\right)}}\overline{A}_I\right]\Phi \nn \\
&-& \frac{3}{2}\kappa a^2\left(\mathcal{H}'-\mathcal{H}^2\right)\sum\limits_I\left(1+\omega_I\right)\zeta_I\,\nn\\
&+&\frac{3}{2}\kappa
a^2\mathcal{H}\sum\limits_I\left[\cfrac{\left(1+3\omega_I\right) \left(1+\omega_I\right)}{a^{3\left(1+\omega_I\right)}}\overline{A}_I\right]\Phi \nn \\
&+&\frac{3}{2}\kappa a^2\mathcal{H}\triangle^{-1}\left[\sum\limits_I\cfrac{\left(1+\omega_I\right)}{a^{3\left(1+\omega_I\right)}}\nabla\left[\delta
A_I\nabla\Phi\right]\right] = 0 \, .
\ea
 Here, the third line represents a sum of summands of the order $\sim\delta A_I\Phi$ (with factor $\sim(1+\omega_I)$), and each summand of this order can be ignored
in view of the fact that the terms of the superior order $\sim\delta A_I$ (with factors $\sim 1$ (see Eq. \rf{a.13}) and $\sim\omega_I$) have already been cancelled in the previous steps. Our reasoning is that
since we keep the terms in our equations up to only a certain order of smallness, previous cancellations of terms do not correspond to identical zero. To respect the adopted accuracy, it is necessary to discard also the terms with higher orders of smallness than these. Therefore, from \rf{a.21} we
obtain
\ba{a.22} &&\frac{1}{2}\kappa a^2\left[\sum\limits_I \cfrac{1+\omega_I}{a^{3\left(1+\omega_I\right)}}\overline{A}_I\right]\sum\limits_I \left(1+\omega_I\right)\zeta_I\,\nn\\
&+&\left(\mathcal{H}'-\mathcal{H}^2\right)\sum\limits_I\left(1+\omega_I\right)\zeta_I = 0 \, ,\ea
and this is an identity. Hence, Eq. \rf{a.9} proves adequate within the considered accuracy.

\subsection{Perfect fluid with nonlinear equation of state}

Let us now turn to the nonlinear perfect fluid with EoS \mbox{$p_J=f_J(\varepsilon_J)$} where $f_J$ is a nonlinear function. In this case, Eq. \rf{2.25} reads
\be{a.23}
\Phi'+\mathcal{H}\Phi=-\frac{\kappa a^2}{2}\sum_J\zeta_J \, ,
\ee
and acting on it with the Laplace operator, we get
\be{a.24}
\triangle\Phi'+\mathcal{H}\triangle\Phi=-\frac{\kappa a^2}{2}\sum_J\triangle\zeta_J \, .
\ee
$\triangle\Phi$ can be expressed with the help of Eq. \rf{2.32}. Differentiating  it with respect to the conformal time $\eta$, we find
\ba{a.25}
\triangle\Phi'&=&\frac{3\kappa a^2}{2}\sum_J\left(\overline F_J+\overline{f_J(F_J)}\right)\Phi'\,\nn\\
&-&\frac{3\kappa a^2 \mathcal{H}}{2}\sum_J\left(\overline F_J+\overline{f_J(F_J)}\right)\Phi+\frac{3\kappa a^2}{2}\sum_J\overline{f_J(F_J)}'\Phi\,\nn\\
&+&\kappa a^2\mathcal{H}\sum_J\delta F_J+\frac{\kappa a^2}{2}\sum_J\delta F'_J\,\nn\\
&-&\frac{3\kappa a^2}{2}\left(\mathcal{H}'+2\mathcal{H}^2\right)\sum_J\zeta_J-\frac{3\kappa a^2\mathcal{H}}{2}\sum_J\zeta'_J\, .
\ea
Then, substituting $\triangle\Phi'$ (from Eq. \rf{a.25}), $\triangle\Phi$ (from Eq. \rf{2.32}) and $\Phi'$ (from \rf{a.23}) into
 Eq. \rf{a.24}, we obtain 
\ba{a.26}
&&\frac{3\kappa a^2 \mathcal{H}}{2}\sum_J\left(\overline F_J+\overline{f_J(F_J)}\right)\Phi\,\nn\\
&-&\frac{3\kappa a^2}{2}\sum_J\left(\overline{\left.\frac{\partial f_J}{\partial \varepsilon_J}\right|_{\varepsilon_J=F_J}F_J'}\right)\Phi-\frac{3\kappa a^2\mathcal{H}}{2}\sum_J\delta F_J\,\nn\\
&-&\frac{\kappa a^2}{2}\sum_J\delta F'_J+\frac{3\kappa a^2}{2}4\mathcal{H}^2\sum_J\zeta_J+\frac{3\kappa a^2\mathcal{H}}{2}\sum_J\zeta'_J\,\nn\\
&=&\frac{\kappa a^2}{2}\sum_J\triangle\zeta_J \, ,
\ea
where we have also employed the equation $\kappa a^2 \sum_J (\ov\varepsilon_J+\ov p_J)=2(\mathcal{H}^2 - \mathcal{H}')$.

To determine the expression for $\zeta_J'$, we now take the gradient of \rf{2.34}, which yields
\ba{a.27}
&{}&\mathbf{B}'\nabla (F_J+f_J(F_J))+\mathbf{B}\left.\left[\left(1+\frac{\partial f_J}{\partial\varepsilon_J}\right|_{\varepsilon_J=F_J}\right)\nabla F'_J\right.\,\nn\\
&+&\left.F'_J\nabla\left(\left.\frac{\partial f_J}{\partial\varepsilon_J}\right|_{\varepsilon_J=F_J}\right)+4\mathcal{H}\nabla(F_J+f_J(F_J))\right] \,\nn\\
&-&\triangle\left(\zeta_J'+4\mathcal{H}\zeta_J+f_J(F_J)\phantom{\frac{\partial f_J}{\partial\varepsilon_J}}\right.\,\nn\\
&-&\left.\left.3\frac{\partial f_J}{\partial\varepsilon_J}\right|_{\varepsilon_J=F_J}\left[F_J+f_J(F_J)\right]\Phi\right)\,\nn\\
&-&\nabla(F_J+f_J(F_J))\nabla\Phi-(F_J+f_J(F_J))\triangle\Phi=0\, ,
\ea
and acting on with the inverse Laplacian, we obtain
\ba{a.28}
\zeta_J'&=&-4\mathcal{H}\zeta_J-f_J(F_J)-3\left.\frac{\partial f_J}{\partial\varepsilon_J}\right|_{\varepsilon_J=F_J}\left[F_J+f_J(F_J)\right]\Phi\,\nn\\
&+&\triangle^{-1}\left[\mathbf{B}'\nabla (F_J+f_J(F_J))+\left.\mathbf{B}\left(1+\frac{\partial f_J}{\partial\varepsilon_J}\right|_{\varepsilon_J=F_J}\right)\nabla F'_J\right.\nn\\
&+&\left.\mathbf{B}F'_J\nabla\left(\left.\frac{\partial f_J}{\partial\varepsilon_J}\right|_{\varepsilon_J=F_J}\right)+4\mathcal{H}\mathbf{B}\nabla(F_J+f_J(F_J))\right]\,\nn\\
&-&\triangle^{-1}\left[\nabla(F_J+f_J(F_J))\nabla\Phi+(F_J+f_J(F_J))\triangle\Phi\right] .
\ea
Substituting  $\zeta_J'$ and $\delta F'_J$ (from equation \rf{2.19}) into \rf{a.26}, after lengthy but simple algebra,
we find
\ba{a.29}
&&\frac{3\kappa a^2}{2}\sum_J\left(\overline{\left.\frac{\partial f_J}{\partial \varepsilon_J}\right|_{\varepsilon_J=F_J}F_J'}\right)\Phi\,\nn\\
&+&\frac{3\kappa a^2\mathcal{H}}{2}\sum_J\left[\overline{f_J(F_J)}+3\left.\frac{\partial f_J}{\partial\varepsilon_J}\right|_{\varepsilon_J=F_J}\left[F_J+f_J(F_J)\right]\Phi\right]\,\nn\\
&-&\frac{3\kappa a^2\mathcal{H}}{2}\sum_J\triangle^{-1}\left[2\mathcal{H}\mathbf{B}\nabla (\delta F_J+\delta f_J)\phantom{\frac{\partial f_J}{\partial\varepsilon_J}}\right.\nn\\
&+&\left.\left.\mathbf{B}\left(1+\frac{\partial f_J}{\partial\varepsilon_J}\right|_{\varepsilon_J=F_J}\right)\nabla \delta F'_J+\mathbf{B}F'_J\nabla\left(\left.\frac{\partial f_J}{\partial\varepsilon_J}\right|_{\varepsilon_J=F_J}\right)\right]\,\nn\\
&+&\frac{3\kappa a^2\mathcal{H}}{2}\sum_J\triangle^{-1}\nabla\left[(\delta F_J+\delta f_J)\nabla\Phi\right]=0 \, ,
\ea
where the terms \mbox{$\propto\kappa a^2\mathcal{H}^2\sum\limits_J\zeta_J$}, \mbox{$\propto\kappa a^2 \mathcal{H}\sum\limits_J\left(\delta F_J+\delta f_J\right)$}, \mbox{$\propto\kappa a^2\sum\limits_J\triangle\zeta_J$}, \mbox{$\propto\kappa a^2 \mathcal{H}\sum\limits_J\left(\overline F_J+\overline{f_J(F_J)}\right)\Phi$} have been cancelled. Back in the representation without $\triangle^{-1}$ (i.e. acting on by $\triangle$), this equation reads
\ba{a.30}
&&\frac{3\kappa a^2}{2}\sum_J\left(\overline{\left.\frac{\partial f_J}{\partial \varepsilon_J}\right|_{\varepsilon_J=F_J}F_J'}\right)\triangle\Phi\,\nn\\
&+&\frac{3\kappa a^2\mathcal{H}}{2}\sum_J\triangle\left[3\left.\frac{\partial f_J}{\partial\varepsilon_J}\right|_{\varepsilon_J=F_J}\left[F_J+f_J(F_J)\right]\Phi\right]\,\nn\\
&-&\frac{3\kappa a^2\mathcal{H}}{2}\sum_J\left[2\mathcal{H}\mathbf{B}\nabla (\delta F_J+\delta f_J)\phantom{\frac{\partial f_J}{\partial\varepsilon_J}}\right.\,\nn\\
&+&\left.\left.\mathbf{B}\left[\left(1+\frac{\partial f_J}{\partial\varepsilon_J}\right|_{\varepsilon_J=F_J}\right)\nabla \delta F'_J+F'_J\nabla\left(\left.\frac{\partial f_J}{\partial\varepsilon_J}\right|_{\varepsilon_J=F_J}\right)\right]\right]\,\nn\\
&+&\frac{3\kappa a^2\mathcal{H}}{2}\sum_J\nabla\left[(\delta F_J+\delta f_J)\nabla\Phi\right]=0 \, .
\ea
The second term  in the first line (i.e. the term in large square brackets) can be expressed with the help of Eq. \rf{2.14}. Then, using equations \rf{2.16} and \rf{2.19}, eliminating the second order terms and exploiting the condition $|\Phi|\ll 1$, we arrive at
\ba{a.31}
&&\sum_J\left(\overline{\left.\frac{\partial f_J}{\partial \varepsilon_J}\right|_{\varepsilon_J=F_J}F_J'}\right)\triangle\Phi-\sum_J\triangle\left[\left.\frac{\partial f_J}{\partial\varepsilon_J}\right|_{\varepsilon_J=F_J}\overline F_J'\Phi\right]\,\nn\\
&-&\mathcal{H}\sum_J\left[2\mathcal{H}\mathbf{B}\nabla (\delta F_J+\delta f_J)\phantom{\frac{\partial f_J}{\partial\varepsilon_J}}\right.\,\nn\\
&+&\left.\left.\mathbf{B}\left[\left(1+\frac{\partial f_J}{\partial\varepsilon_J}\right|_{\varepsilon_J=F_J}\right)\nabla \delta F'_J+F'_J\nabla\left(\left.\frac{\partial f_J}{\partial\varepsilon_J}\right|_{\varepsilon_J=F_J}\right)\right]\right]\,\nn\\
&+&\mathcal{H}\sum_J\nabla\left[(\delta F_J+\delta f_J)\nabla\Phi\right]=0 \, .
\ea
Now, to compare some of terms above with the previously cancelled ones, we should remember that, first, all terms in \rf{a.31} were divided by $\kappa a^2$ and second, the expression was acted on with $\triangle$. This means that either we should apply $\triangle$ to the cancelled terms or, vice versa, apply $\triangle^{-1}$ to the remaining ones. For example, the cancelled term $\propto\kappa a^2\mathcal{H}^2\sum\limits_J\zeta_J \; \Rightarrow\; \mathcal{H}^2\sum_J\triangle\zeta_J=\mathcal{H}^2\sum_J\nabla \left[\left(F_J+f_J(F_J)\right)\tilde{\mathbf{v}}_J\right]\sim \mathcal{H}^2\sum_J\nabla(F_J\tilde{\mathbf{v}}_J) $, where we have used \rf{2.23}. Then, employing the conditions $\delta F_J\sim \delta f_J, |\partial f_J/\partial\varepsilon_J|\sim 1$ and taking into account that time derivatives with respect to $\eta$ are proportional in order to the Hubble parameter $\sim \mathcal{H}$, it can be deduced that all three terms in the second line behave as $\mathcal{H}^2\sum_J\nabla ({\bf{B}}\delta F_J)$. 
Thus, the ratio of these terms and the cancelled term is of the order of $(\delta F_J/F_J)(B/\tilde{v}_J)\sim B\tilde{v}_J/\Phi\sim \epsilon$ (see the estimate \rf{2.18}). Similarly, the very last term $\propto \mathcal{H}\delta F_J\Phi$ may also be neglected provided that terms of higher order, i.e. \mbox{$\propto\kappa a^2\mathcal{H}\sum_J(\delta F_J+\delta f_J)$} have already been cancelled in the above steps. Eventually, \rf{a.31} is reduced to
\ba{a.32}
&{}&\triangle\left[\Phi\sum_J\left(\overline{\left.\frac{\partial f_J}{\partial \varepsilon_J}\right|_{\varepsilon_J=F_J}\left[-3\mathcal{H}\left(F_J+{f_J(F_J)}\right)\right]}\right.\right.\,\nn\\
&+&\left.\left.3\mathcal{H}\left( F_J+{f_J(F_J)}\right)\left.\frac{\partial f_J}{\partial\varepsilon_J}\right|_{\varepsilon_J=F_J}\right.\right.\,\nn\\
&-&\left.\left.3\mathcal{H}\left( \delta F_J+\delta {f_J(F_J)}\right)\left.\frac{\partial f_J}{\partial\varepsilon_J}\right|_{\varepsilon_J=F_J}\right)\right]=0\, ,
\ea
and taking into consideration that $\left.\frac{\partial f_J}{\partial\varepsilon_J}\right|_{\varepsilon_J=F_J}$ is of the order of unity, we are left with the term $\propto \Phi\mathcal{H}\left(\delta F_J+\delta f_J(F_J)\right)$, which is again to be neglected in comparison to previously cancelled terms in the expression. Therefore, the LHS of this equation is equal to zero up to adopted accuracy, and this serves as the proof of the equation \rf{a.24}.

To conclude this appendix, it is worth noting that in a similar way we can also prove the relation ${\bf{B}}'+2\mathcal{H}{\bf{B}}=0$. 


\end{document}